# Impedance Design of Excitation Lines in Adiabatic Quantum-Flux-Parametron Logic Using InductEx

Naoki Takeuchi, *Member, IEEE*, Hideo Suzuki, Coenrad J. Fourie, *Senior Member, IEEE*, and Nobuyuki Yoshikawa, *Senior Member, IEEE*

*Abstract*—The adiabatic quantum-flux-parametron (AQFP) is an energy-efficient superconductor logic family that utilizes adiabatic switching. AQFP gates are powered and clocked by ac excitation current; thus, to operate AQFP circuits at high clock frequencies, it is required to carefully design the characteristic impedance of excitation lines (especially, above AQFP gates) so that the microwave excitation current can propagate without reflections in the entire circuit. In the present study, we design the characteristic impedance of the excitation line using InductEx, which is a three-dimensional parameter extractor for superconductor devices. We adjust the width of an excitation line using InductEx such that the characteristic impedance is maintained at 50 Ω even above an AQFP gate. Then, we fabricate test circuits to verify the impedance of the excitation line. We measure the impedance using time domain reflectometry (TDR). We also measure the S parameters of the excitation line to investigate the maximum possible clock frequency. Our experimental results indicate that the characteristic impedance of the excitation line agrees well with the design value even above AQFP gates, and that clock frequencies beyond 5 GHz are possible in large-scale AQFP circuits.

## I. Introduction

SUPERCONDUCTOR logic families [1]–[4] are crucial for achieving cryogenic information systems, such as energy-efficient microprocessors [5], [6], single-photon image sensors [7], [8], and quantum computers [9], [10]. We have been investigating adiabatic quantum-flux-parametron (AQFP) logic [11], which is an adiabatic logic family based on the quantum flux parametron (QFP) [12], [13]. AQFP circuits can operate with an energy dissipation of approximately $10^{-21}$ J per Josephson junction [14] by using underdamped high-critical-current-density Josephson junctions, which maximize the benefit of adiabatic switching [15], [16]. We have designed and demonstrated various AQFP circuits, such as a microprocessor [6], a cryogenic detector interface [8], and a stochastic problem solver [17].

To conduct adiabatic switching, the potential energy of an AQFP gate is modulated between a single-well shape and a double-well shape by applying an ac excitation current. This means that all AQFP gates, including both combinational and sequential circuits, must be powered and clocked by the excitation current. Furthermore, unlike complementary metal–oxide–semiconductor (CMOS) logic, AQFP gates must be clocked in the order of logic operation with moderate clock skews. We have proposed clocking schemes for AQFP logic [18], [19] to appropriately distribute the excitation current in the entire AQFP circuit, whereby we have demonstrated small-to-medium-scale AQFP circuits at GHz-range clock frequencies [14], [19]. To apply the above clock schemes to large-scale AQFP circuits at high clock frequencies, it is required to carefully design the characteristic impedance of excitation lines (i.e., clock paths) so that the microwave excitation current is applied equally to each AQFP gate without standing waves. However, it has been difficult to design the impedance of excitation lines precisely because the physical structure of excitation lines is too complicated to allow theoretical estimation of the impedance; a part of an excitation line is placed above AQFP gates and thus cannot be treated as a simple microstrip line, as will be shown later. Note that the impedance of a simple microstrip line can be analytically estimated [20], [21], and that microstrip and strip lines are used for interconnection in RSFQ circuits [22].

In this paper, we precisely design the impedance of excitation lines using InductEx [23], which is a three-dimensional inductance extractor for superconductor devices. Recently, characteristic impedance calculation capability was added to InductEx [24], which simultaneously calculates the magnetoquasistatic inductance, with the inclusion of kinetic inductance, and the electroquasistatic capacitance, with the inclusion of multiple dielectric layers. We adjust the width of an excitation line using InductEx such that the characteristic impedance is maintained at 50 Ω even above an AQFP gate. Then, we fabricate test circuits to verify the impedance of the designed excitation line. We measure the impedance using time domain reflectometry (TDR) [25]. We also measure the S parameters of the excitation line to investigate the maximum possible clock frequency.

This work was supported by KAKENHI (No. 18H01493 and No. 19H05614) from the Japan Society for the Promotion of Science (JSPS). *(Corresponding author: Naoki Takeuchi)*.

N. Takeuchi and H. Suzuki are with the Institute of Advanced Sciences, Yokohama National University, 79-5 Tokiwadai, Hodogaya, Yokohama 240-8501, Japan (e-mail: takeuchi-naoki-kx@ynu.ac.jp; suzuh@ynu.ac.jp).

C. J. Fourie is with the Department of Electrical and Electronic Engineering, Stellenbosch University, Stellenbosch, South Africa (e-mail: coenrad@sun.ac.za).

N. Yoshikawa is with the Institute of Advanced Sciences, Yokohama National University, 79-5 Tokiwadai, Hodogaya, Yokohama 240-8501, Japan; and also with the Department of Electrical and Computer Engineering, Yokohama National University, 79-5 Tokiwadai, Hodogaya, Yokohama 240-8501, Japan (e-mail: nyoshi@ynu.ac.jp).

## II. IMPEDANCE DESIGN

Figure 1(a) illustrates an AQFP buffer, which is one of the most fundamental logic gates in AQFP logic, and Fig. 1(b) shows typical waveforms for the buffer. The buffer is in the form of an rf superconducting quantum interference device (SQUID) whose Josephson junction is replaced with a dc SQUID (J1-$L_1$-$L_2$-J2). The excitation current $I_x$ modulates the critical current of the dc-SQUID, i.e., the potential energy shape of the buffer. A dc offset current $I_d$ applies an offset flux of $0.5\Phi_0$ for four-phase clocking [18], where $\Phi_0$ is the flux quantum. While $I_x$ increases, either J1 or J2 switches depending on the polarity of the input current $I_{in}$. Consequently, the state current $I_{st}$, which represents the logic state of the buffer, appears through the load inductor $L_q$. A positive $I_{st}$ represents a logic 1, whereas a negative $I_{st}$ represents a logic 0. A signal transformer composed of $L_q$ and $L_{out}$ is included to facilitate logic negation; while the AQFP gate shown in Fig. 1(a) operates as a buffer for a positive $k_{out}$, it operates as an inverter for a negative $k_{out}$.

Since an AQFP gate is powered and clocked by $I_x$ as shown in Fig. 1, the operating frequency (clock frequency) of an AQFP gate is equal to the frequency of $I_x$. Therefore, the characteristic impedance $Z_0$ of the excitation line $L_x$ should be carefully designed such that $I_x$ can propagate through $L_x$ without reflections even at GHz-range clock frequencies. Figure 2(a) shows the physical layout of an AQFP buffer designed for the AIST 10 kA/cm$^2$ Nb high-speed standard process (HSTP) [18], whose dimensions are 20 μm (width) by 40 μm (height). The HSTP includes four Nb layers (M1 through M4), where M1 is used for ground planes and Josephson junctions are formed between M2 and M3. Here, we calculate $Z_0$ using InductEx and adjust the line width of $L_x$ such that $Z_0$ becomes 50 Ω. One difficulty in impedance calculation using InductEx is that the current version of InductEx cannot handle non-planarized processes such as HSTP, due to modeling limitations when the surface model for capacitance calculation is generated. As shown in Fig. 2(a), part of $L_x$ (in M4) is placed above inductors $L_1$ and $L_2$ (in M3); thus, the distance $d$ between $L_x$ and the ground plane (in M1) is different for segments A and B, which cannot be handled by the current InductEx. Therefore, we calculate the impedance (i.e., inductance and capacitance) of $L_x$ in segment A and that in segment B separately, and then calculate $Z_0$ from the total inductance and capacitance in segments A and B. Figure 2(b) shows the physical model used to calculate the impedance in segment A, where $L_x$ is a microstrip line placed above $L_1$ and $L_2$ with $d = 1.6$ μm. The Josephson junctions (J1 and J2) are replaced with ground contacts (GC1 and GC2) because $L_1$ and $L_2$ are connected to the ground plane via J1 and J2. InductEx yields the inductance between ports P1 and P2 as well as the capacitance between the microstrip line $L_x$ and ground. Figure 2(c) shows the physical model for segment B, where $L_x$ is a simple microstrip line with $d = 1.2$ μm. Using the physical models shown in Figs. 2(b) and (c), we calculate $Z_0$ and adjust the line width of $L_x$, where the relative permittivity of SiO$_2$ layers between Nb layers is set to 3.9. We found that $Z_0$ becomes approximately 50 Ω when the line width is 1.5 μm; the inductance and capacitance with regard to $L_x$ in segment A are 5.10 pH and 2.33 fF, respectively, and those in segment B are 1.80 pH and 0.501 fF, respectively, which results in a $Z_0$ of 49.4 Ω and a phase velocity $v_p$ of 1.43×10$^8$ m/s. It should be noted that the resulting line width is narrower than that of a simple 50-Ω microstrip line with $d = 1.2$ μm (2.4 μm) and that the phase velocity is lower than that of a simple microstrip line with $d = 1.2$ μm (1.63×10$^8$ m/s). This is because $L_1$ and $L_2$ are connected to the ground plane, so that the capacitance between $L_x$ and the ground plane increases in segment A. Meanwhile, inductance decreases slightly in segment A; the inductance per unit length in segment A is 0.34 pH/μm, whereas that in segment B is 0.36 pH/μm.

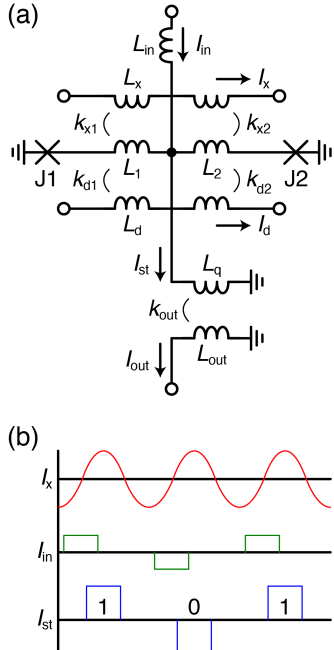

Fig. 1. AQFP buffer. (a) Schematic. $I_x$ powers and clocks the buffer. (b) Typical waveform. The logic state $I_{st}$ is determined by the polarity of $I_{in}$.

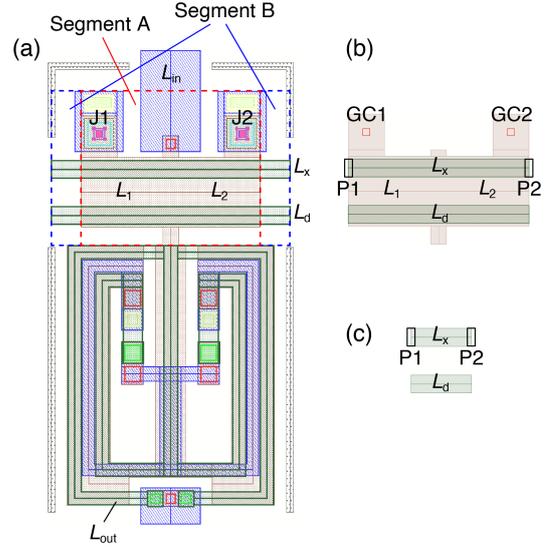

Fig. 2. Layout design. (a) AQFP buffer. (b) Physical model for segment A. (c) Physical model for segment B.

## III. Experiments

We measure $Z_0$ to verify the calculation results obtained using InductEx. Figure 3(a) illustrates the experimental setup used to measure $Z_0$ through TDR. An array of AQFP buffers are coupled to a long meandering excitation line, which is represented by a transmission line rather than lumped inductors. A step generator (Tektronix, SD-24) applies an incident pulse with a short rise time to the excitation line, and then reflected pulses appear where the characteristic impedance changes. A sampling oscilloscope (Tektronix, 11801) samples the resultant wave of the incident and reflected pulses, thereby measuring $Z_0$. It should be noted that a large reflected pulse appears at the end of the excitation line because of the open end. This makes it easy to measure both $Z_0$ and $v_p$, as will be shown later. We also measure the S parameters of the excitation line to determine the maximum operating frequency. Figure 3(b) illustrates the experimental setup used to measure the S parameters. As with Fig. 3(a), an array of AQFP buffers are coupled to a long meandering excitation line. Both input and output ports are terminated by a vector network analyzer (VNA) (Keysight, P9374A). The VNA measures the transmission loss ($S_{21}$) and reflection loss ($S_{11}$) of the excitation line. Figure 4 shows a micrograph of the chip including the circuits under test (CUTs) for the TDR and S parameter measurements. This chip was fabricated through HSTP with a 5×5 mm die, including five CUTs (CUT 1 through CUT 5). CUT 1 through CUT 3 are for the TDR measurement and correspond to the schematic in Fig. 3(a). The length of the excitation line ($l$) varies between CUT 1 through CUT 3: $l$ = 15 mm, 30 mm, and 45 mm for CUT 1 through CUT 3, respectively. CUT 4 is a through line used as a reference in the S parameter measurement. CUT 5 is for the S parameter measurement and corresponds to the schematic in Fig. 3(b) with $l$ = 15 mm. The inset shows part of an AQFP buffer array, where the physical layout of each buffer is the same as that shown in Fig. 2(a).

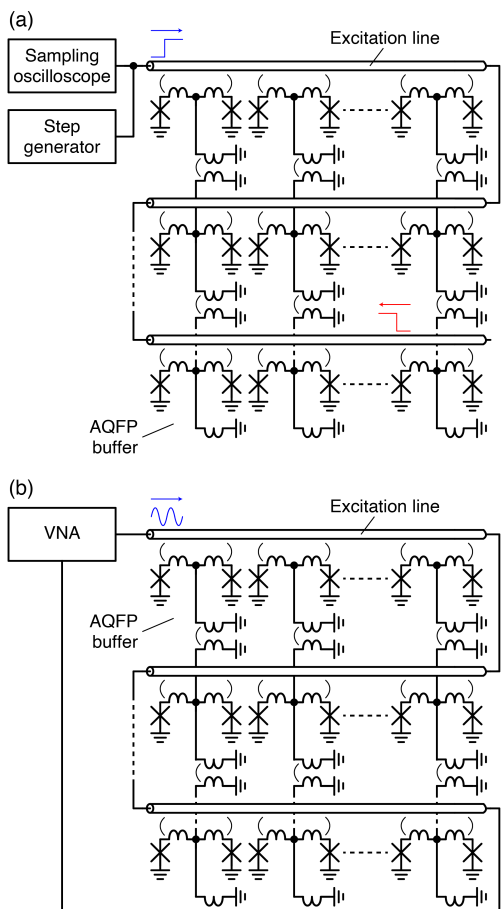

Fig. 3. Experimental setups using an array of AQFP buffers for (a) TDR and (b) S parameter measurement.

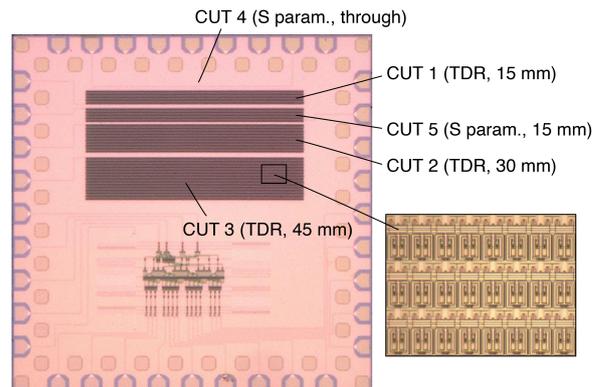

Fig. 4. Micrograph of the chip including CUTs. CUT 1 through CUT 3 are for TDR. CUT 4 and CUT 5 are for the S parameter measurement.

We conducted experiments using a wide-bandwidth cryoprobe at 4.2 K in liquid He. Figure 5(a) shows the TDR measurement results for CUT 1 through CUT 3. The blue, green, and red solid lines are the voltage signals for, respectively, CUT 1, CUT 2, and CUT 3. The amplitude of the incident pulse ($V_0$ = 0.250 V) corresponds to 50 Ω, and the voltage difference $\Delta V$ from $V_0$ represents the characteristic impedance of the CUTs. The sharp rises at the edges of the voltage signals represent the reflected pulses at the open ends, and the fluctuations between 85.8 ns and 86.0 ns represent impedance changes due to the pads on the chip. Thus, the voltage signal between 86.0 ps and the rise at the edge represent the $Z_0$ of the excitation line. Figure 5(a) shows that $\Delta V$ for the excitation line is 6.49 mV, which is an average between 86.24 ns and 86.44 ns for CUT 3. Therefore, $Z_0 = (1 + \rho)/(1 - \rho) \times 50 = 52.7$ Ω, where $\rho = \Delta V/V_0 = 0.0260$. The measured $Z_0$ agrees well with the simulation result (49.4 Ω). The slight discrepancy may have arisen because the impedance calculation using InductEx does not take into account the shrinkage in the line width of $L_x$ (typically, approximately 0.1 μm). We also measured $v_p$ based on Fig. 5(a). The difference in the timing of the rising edges in the voltage signals ($t_1$ through $t_3$) is attributed to the difference in the excitation line length ($\Delta l$); thus, this time difference corresponds to $2\Delta l/v_p$, where the coefficient 2 is for the round-trip length. Figure 5(b) shows the timing of the rising edges as a function of the round-trip length of an excitation line ($2l$), where the solid line represents the linear regression given by $\alpha \times 2l + \beta$ ($\alpha$ = 7.38 ps/mm, $\beta$ = 86.0 ps). Consequently, $v_p$ is given by $1/\alpha = 1.35 \times 10^8$ m/s, which agrees well with the simulation result ($1.43 \times 10^8$ m/s).

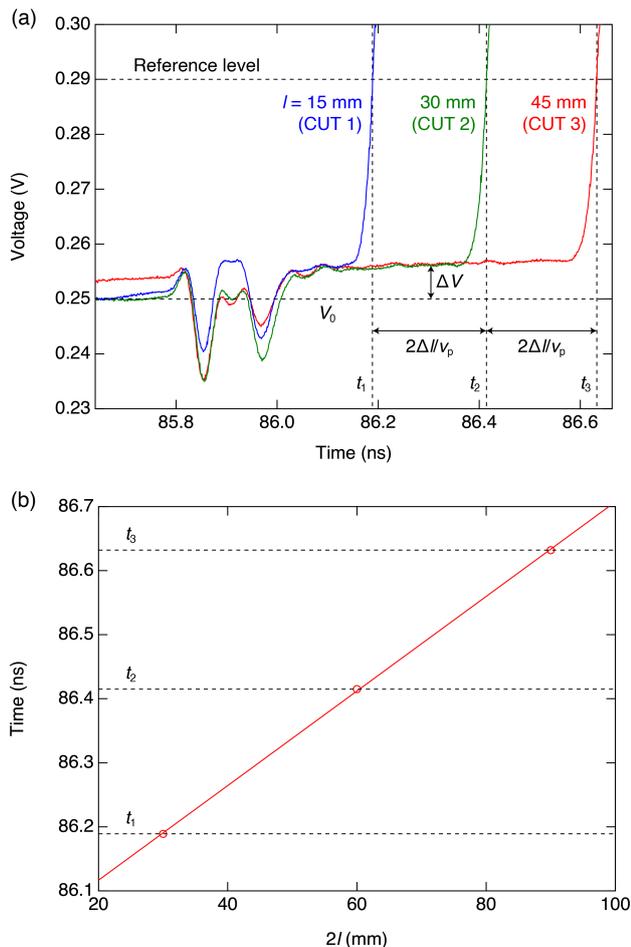

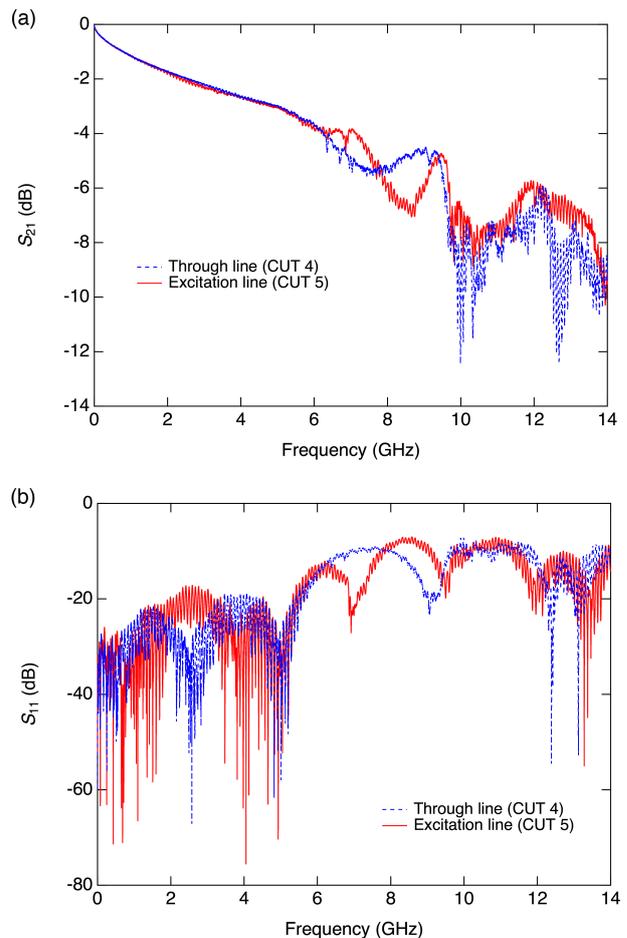

Fig. 5. Measurement results for CUT 1 through CUT 3. (a) TDR. (b) Timing of rising edges in TDR as a function of the excitation line length.

Fig. 6. Measurement results for CUT 4 and CUT 5. (a) Transmission loss and (b) reflection loss.

Figure 6 shows the measurement results of $S_{21}$ and $S_{11}$ for CUT 4 and CUT 5. The blue dashed lines are the results for CUT 4, and the red solid lines are for CUT 5. Since CUT 4 is a through line, the S parameters for CUT 4 represent the frequency response of the cryoprobe used in the experiment. Figure 6(a) shows that $S_{21}$ for CUT 5 is almost the same as that for CUT 4 up to approximately 6 GHz. Figs. 6(a) and (b) show that for CUT 5 there is no clear dip in $S_{21}$ or peak in $S_{11}$ at the resonant frequency ($v_p/2l$ = 4.5 GHz). The above measurement results indicate that a microwave excitation current can propagate along a long excitation line above AQFP buffers without significant reflections, and that high clock frequencies are possible in large-scale AQFP circuits. The measurement results also validate piecewise impedance calculation with InductEx, and the applicability of quasistatic calculation of characteristic impedance in superconductor integrated circuit layouts at frequencies far below the gap frequency.

## IV. CONCLUSION

We designed the characteristic impedance of the excitation line in an AQFP buffer using InductEx so that the microwave excitation current can propagate in a large-scale AQFP circuit without reflections. We verified that the characteristic impedance of the excitation line agrees well with the design value in the TDR experiment. We also found that a microwave excitation current can propagate along a long excitation line without significant reflections in the S parameter measurement. It is noteworthy that the excitation line design for the buffer is applicable to other logic gates because in AQFP logic the physical layout design of all logic gates is based on that of a buffer [26]. Our measurement results demonstrated that a microwave excitation current can propagate appropriately in a large-scale AQFP circuit with the help of impedance design using InductEx.

ACKNOWLEDGMENT

The circuits were fabricated in the Clean Room for Analog-digital superconductiVITY (CRAVITY) of the National Institute of Advanced Industrial Science and Technology (AIST). The authors thank Kyle Jackman for his valuable contributions.